\documentstyle[11pt]{article}
\input epsf

\title{\bf Inclusive cross-sections for jet production in
nucleus-nucleus collisions in perturbative QCD}
\author{M.A.Braun\\
Dep. of High Energy physics,
 University of S.Petersburg,\\
198504 S.Petersburg, Russia}
\def\beq{\begin{equation}}
\def\eeq{\end{equation}}
\def\noi{\noindent}
\begin{document}
\maketitle
\medskip
\noi{\bf Abstract.} Inclusive cross-sections for gluon jet
production are studied numerically in the perturbative QCD pomeron
model for central collisions of identical nuclei at high energies.
Two forms for the inclusive cross-sections, with and without
emission from the triple pomeron vertex, are compared. The
difference was found to reduce to a numerical factor $0.8\div 0.9$
for momenta below the saturation momentum $Q_s$. Above $Q_s$ no
difference was found at all. The gluon spectrum was found to be
$\sim A$ at momenta $k$ below $Q_s$ and $\sim A^{1.1}$ above it. At
large  $k$ the spectrum goes like $1/k^{2.7\div 3.3}$ flattening
with energy. The multiplicities turned out to be proportional to $A$
with a good precision. Their absolute values are high and grow
rapidly with energy in accordance with the high value of the BFKL
intercept. \vspace{1.5cm}

\section{Introduction}
In view of the new experimental data on heavy-ion collisions  at RHIC and
future such data to be obtained at LHC one would like to have
predictions for the spectra of produced secondaries based on the
fundamental theory and not purely phenomenological. At present
the only candidate for this is the hard pomeron model derived
from perturbative QCD. Originally constructed for the description
of high-energy low -$x$ hadronic scattering  (the BFKL model~\cite{BFKL})
it has subsequently been generalized to hadronic or deep inelastic scattering
on nuclei ~\cite{kov,bra1} and nucleus-nucleus scattering ~\cite{bra2}.
The model suffers from a
serious drawback related to the use of fixed and not running
strong coupling constant. Curing it does not look too promising,
since due to absence of ordering of momenta in the model, it also
means solving the confinement problem. However in spite of this
defect the model seems to describe high-energy phenomena
in a qualitatively reasonable manner. Also  attempts to include
the running of the coupling in some effective way have shown that the
effect of the running is not at all overwhelming, although introduces
some quantitative changes into the predictions. So, also for lack of something
better, the perturbative QCD pomeron model appears to give a reasonable
basis for the discussion of particle production in high-energy heavy-ion
collisions. Of course due to the  perturbative character of the model it
can only give predictions for production of jets, leaving jet-to-hadrons conversion
to non-perturbative fragmentation mechanism.

From the start it has to be stressed that  heavy-ion collision amplitudes
are described in the model  by complicated equations, whose solution is
quite difficult to obtain even numerically (see ~\cite{bra3} for partial results).
Happily, as was shown in ~\cite{bra4}, due to Abramovsky-Gribov-Kancheli (AGK)
cancellations  ~\cite{AGK}, to find the single inclusive distributions one does
 not have to
solve these equations, but only to sum the appropriate sets of fan diagrams,
which is accomplished by the non-linear evolution equation of ~\cite{kov,bra1}.
Still this operation involves a numerical study of considerable complexity.
So up to now there has been no consistent calculation of the jet spectra
for realistic nuclei, although some preliminary attempts has been done
in ~\cite{nestor,kov2, KNV,KLN}. In all cases however the authors relied on very drastic
simplifications
from the start choosing for the nuclear structure and/or for the gluon
distributions in the colliding  nuclei some primitive explicit forms
 in accordance with their own taste and prejudice. In fact these
forms appear to be  rather far from realistic ones, which correspond to
actual participants and follow from the calculations.
This gave us motivation to calculate numerically the jet spectra in heavy ion
collisions as predicted by the hard pomeron model in a consistent manner.

Another goal of the present calculations has been to compare the
results obtained on the basis of the expression for the inclusive
cross-section which follows from the AGK rules applied to the
diagrams with QCD pomerons interacting via the three-pomeron
coupling ~\cite{bra4} with a somewhat different expression obtained
from the colour dipole picture ~\cite{KT}. Our calculations show
that these two formally different expressions lead to completely
identical results at momenta of the order or higher than the value
of the so-called saturation momentum $Q_s$. At momenta substantially
lower than $Q_s$ the colour dipole cross-sections differ from the
ones from the AGK rules by a universal constant factor $\sim 0.8\div
0.9$.

In both cases the  spectra at
momenta below  $Q_s$ are found to be proportional to the
number of participants ($\propto A$ for collisions of identical nuclei)
 and not to the number of collisions($\propto A^{4/3}$). Since $Q_s$
grows with energy very fast, the region where the spectra are $\propto A$
extends with energy to include all momenta of interest. At momenta
greater than $Q_s$ the spectra grow faster than $A$ but still much slowlier
than $A^{4/3}$ (a numerical fit gives something like $\propto A^{1.1}$).

Note that in the last years a few more phenomenologically oriented
studies of particle production in nucleus-nucleus production have
been presented, in the framework of the color-condensate model
~\cite{ML} solved in the classical approximation on the lattice
~\cite{KNV} and in the saturation model ~\cite{KLN}.
 In both
approaches quantum evolution of the nuclear gluon density was neglected
and the saturation momentum was introduced as a parameter fitted to
the experimental data at RHIC. Although
 some of their predictions (proportionality
of the multiplicity to $A$ modulo logarithms) agree with our calculations
with full quantum evolutions, the quantitative results are rather different.
We postpone a more detailed discussion of this point until our Conclusions.

\section{Basic equations}
Our basic quantity will be the inclusive cross-section $I_{AB}(y,k)$
to produce a jet with the transverse momentum
$k$ at rapidity $y$ in a collision of two nuclei with atomic numbers
$A$ and $B$:
\beq
I_{AB}(y,k)=\frac{(2\pi)^2d\sigma}{dyd^2k}.
\eeq
It can be represented as an integral over the impact parameter $b$:
\beq
I_{AB}(y,k)=\int d^2bI_{AB}(y,k,b).
\eeq
Our study will be restricted to the inclusive cross-sections
at fixed impact parameter $b=0$ (central collisions).
 We shall also limit ourselves to collisions of identical
nuclei $A=B$  and for brevity denote $I_{AA}\equiv I_A$ and so on.
The corresponding multiplicity at fixed rapidity $y$ and $b=0$
will be given by
\beq
\mu_{A}(y)=\frac{1}{\sigma_{A}(b=0)}
\int\frac{d^2k}{(2\pi)^2}I_{A}(y,k,b=0),
\eeq
where $\sigma_{A}(b)$ is the total inelastic cross-section for the
collision of two identical nuclei at fixed impact parameter $b$.
For heavy nuclei one expects that $\sigma_A(b=0)\simeq 1$,
so that the multiplicity is just the integral of the inclusive cross-section
 over the momenta.

As argued in ~\cite{bra2}, in the perturbative QCD with a large number of colours
the nucleus-nucleus interaction is described by a set of tree diagrams
constructed with BFKL pomeron Green functions and triple pomeron vertices
for their splitting and fusing. The structure of the interaction at the
vertex is illustrated in Fig. 1, in which horizontal lines correspond to
real gluons produced in the intermediate states and vertical and inclined  lines
describe  propagating reggeized gluons. From this structure one sees that
the produced gluons are contained in the intermediate states of the interacting
pomerons, so that to get the inclusive cross-section one has to "open" these
pomerons, that is to fix the momentum of one of the intermediate real gluons
in them. A similar production mechanism in the old-fashioned local pomeron
model was proven to lead to the inclusive cross-section given by
a convolution of two sets of fan diagrams connecting
the emitted particle to the two nuclei times the vertex for
the emission (Fig. 2$a$). The proof was based on the
AGK rules  appropriately adjusted for the
triple pomeron interaction ~\cite{cia}.
It was later shown  in ~\cite{BW} that the AGK rules
are fulfilled for interacting BFKL pomerons.
So the same arguments as in ~\cite{cia}
allow to demonstrate that for the collision of two nuclei
the inclusive cross-section will
be given by  the same Fig.2$a$, that is, apart from the emission
vertex,  by the convolution of two sums of fan diagrams,
constructed of BFKL pomerons and triple pomeron verteces, propagating
from the emitted particle towards the two nuclei ~\cite{bra4}.

Taking into account the form of the emission vertex (see ~\cite{bra4})
 we obtain in our case
($A=B$ and fixed $b=0$)
\beq
I_{A}(y,k)=\frac{8N_c\alpha_s}{k^2}\int d^2\beta
d^2re^{ikr}[\Delta\Phi_A(Y-y,r,\beta)]
[\Delta\Phi_A(y,r,\beta)],
\eeq
Here  $\Phi(y,r,\beta)$ is the sum
of all fan diagrams connecting the
pomeron at rapidity $y$ and of the transverse dimension $r$ with the
colliding nuclei at distance $\beta$ from their centers. One of the nuclei
is assumed to be at rest  and the other at the overall rapidity $Y$.
$\Delta$'s are the two-dimensional Laplacians applied to $\Phi$'s.

Later from the colour dipole formalism a slightly different form for
the inclusive cross-section was derived in ~\cite{KT}. For the
dipole-nucleus scattering case it corresponds to changing \beq
2\Phi_A(y,\beta,r)\to 2\Phi_A(y,\beta,r)-\Phi^2_A(y,\beta,r). \eeq
Note that in ~\cite{KT} it was erroneously stated that the change
was from the "quark dipole"  $\Phi$ to the "gluon dipole"
$2\Phi-\Phi^2$. As seen from ( 5 ) it is not. In fact the change is
equivalent to adding to the AGK contribution ( 4 ) a new one which
has the meaning of the emission of the gluon from the triple pomeron
vertex itself. Such a contribution is not prohibited in principle.
From our point of view, taking into account the structure of the
vertex shown in Fig. 1, its appearance is difficult to understand.
However in this paper we do not pretend to discuss the validity of
the two proposed formulas for the inclusive cross-sections on the
fundamental level. Rather we shall compare the cross-sections which
follow from them after numerical calculations.

For the nucleus-nucleus case the recipe of ~\cite{KT} implies taking into account two
new diagrams for the inclusive cross-sections shown in Fig. 2 $b$ and $c$.
As a result one finds, instead of ( 4 ), the Kovchegov-Tuchin (KT) cross-section
\[
I_{A}^{KT}(y,k)=\frac{4N_c\alpha_s}{k^2}\int d^2\beta
d^2re^{ikr}\Big[2\Delta\Phi_A(Y-y,r,\beta)\Delta\Phi_A(y,r,\beta)-\]\beq
\Delta\Phi_A(Y-y,r,\beta)\Delta\Phi_A^2(y,r,\beta)-,
\Delta\Phi_A^2(Y-y,r,\beta)\Delta\Phi_A(y,r,\beta)\Big].
\eeq

Function $\phi_A(y,r,\beta)=\Phi_A(y,r,\beta)/(2\pi r^2)$, in the momentum space,
 satisfies the well-known
non-linear equation ~\cite{kov,bra1}
\beq
\frac{\partial\phi_A(y,q,\beta)}{\partial \bar{y}}=-H\phi_A(y,q,\beta)-\phi_A^2 (y,q,\beta),
\eeq
where $\bar{y}=\bar{\alpha}y$, $\bar{\alpha}=\alpha_sN_c/\pi$,
$\alpha_s$ and $N_c$ are the strong coupling constant and the number
of colours,
respectively, and $H$ is the BFKL Hamiltonian. Eq. ( 7 ) has to be solved
with an initial condition at $y=0$ determined by the colour dipole
distribution in the nucleon smeared by the profile function of the
nucleus.
Both cross-sections ( 4 ) and ( 6 ) can be expressed via function
\beq
h_A(y,q,\beta)=q^2\nabla^2_q\phi_A(y,q,\beta),
\eeq
which has the meaning of internal gluon density in each of the colliding nuclei.
One easily obtains for ( 4 )
\beq
I_{A}(y,k)=\frac{8N_c\alpha_s}{k^2}\int d^2\beta
d^2qh_A(Y-y,k-q,\beta)h_A(y,q,\beta).
\eeq
For ( 6 ) one also obtains a factorized expression similar to  ( 9 )
\beq
I_{A}^{KT}(y,k)=\frac{8N_c\alpha_s}{k^2}\int d^2\beta
d^2qh_A(Y-y,k-q,\beta)\Big[w_A(y,q,\beta)-h_A(y,q,\beta)\Big],
\eeq
where $w_A(y,q,\beta)$ is a new function, which however can be expressed
via $h_A$:
\beq
w_A(y,k,\beta)=\frac{k^2}{2\pi}\int \frac{d^2q}{q^2(k-q)^2}h_A(y,k-q,\beta)h_A(y,q,\beta).
\eeq

Function $h_A(y,k,\beta)$ has a normalization property ~\cite{bra4}
\beq
 \int \frac{d^2k}{k^2}h_A(y,k,\beta)=1 \eeq and at sufficiently
high $y$  acquires a scaling property \beq
h_A(y,k\beta)=h\Big(k/Q_s(y,\beta)\Big),
\eeq
where $Q_s(y,\beta)$
is the above-mentioned saturation momentum. From ( 12 ) and ( 13 )
one easily establishes some properties of the new function $w_A$.
Obviously it scales with the same saturation momentum when $h_A$
does
\beq
 w_A(y,k\beta)=w\Big(k/Q_s(y,\beta)\Big).
\eeq
 At
$k\to\infty$ it has the asymptotic
 \beq
w_A(y,k,\beta)_{k\to\infty}\sim 2h_A(y,k,\beta)
\eeq
 and finally
\beq
\int d^2k w_A(y,k,\beta)=2\int d^2k h_A(y,k,\beta).
\eeq

These properties immediately allow to make some preliminary comparison
between the cross-sections given by ( 4 ) and ( 6 ). Obviously if $k/Q_s$ is large
both expressions give the same cross-section due to ( 16 ). In the opposite
limit of small $k/Q_s$, the scaling property allows to conclude that the ratio of the two
cross-sections is a universal constant which does not depend on $y$, nor on $A$
nor on $\beta$. Our numerical results  confirm these predictions.

\section{Results}

In our study we have taken the initial condition in accordance with
the Golec-Biernat distribution ~\cite{gobi}, duly generalized for the nucleus:
\beq
\phi_A(0,q,\beta)=-\frac{1}{2}a_A(\beta)\,{\rm Ei}
\left(-\frac{q^2}{0.218\, {\rm GeV}^2}\right),
\eeq
with
\beq
a_A=20.8\, {\rm mb}\,AT_A(\beta).
\eeq
where $T_A(\beta)$ is the standard nuclear profile function, which we have
taken from the Woods-Saxon nuclear density.

Evolving $\phi_A(y,q,\beta)$ up to values ${\bar y}=8$ we found the
inclusive cross-sections ( 4 )and ( 6 ) at center rapidity  for
energies corresponding to the overall rapidity
$Y=\bar{Y}/\bar{\alpha}$. with $\bar{Y}=16$. Taking $\alpha_s=0.2$
this gives $Y\sim 80$. This value  is is far beyond the present
possibilities and was chosen only to follow the asymptotical
behavior of the cross-sections at super-high energies and compare it
with the situation at energies  available presently or the near
future. The overall cutoffs for integration momenta in Eq.( 7 ) were
taken according to $1.10^{-16}$ GeV/c  $< q < 1.10^{+16}$ GeV/c. For
the (fixed) value of the strong coupling constant we have taken
$\alpha_s=0.2$

We first discuss the cross-sections ( 4 ) obtained from the AGK rules.
They are illustrated in Figs. 3-5.
To see the absolute values of the inclusive cross-sections at
different energies, in Fig. 3 we present
them  for $A=9$, $\bar{Y}=4,8, 16$  and $y=Y/2$.
To illustrate the change of their form  with energy,
in Fig.4 we present the same distributions at $k>0.3$ GeV/c
normalized to unity and multiplied
by $k^2$ to
exclude the trivial $1/k^2$ dependence present in ( 4 ),
\beq
J_{9}(y,k)=k^2I_{9}(y,k)/\int \frac{d^2k'}{(2\pi)^2}I_{9}(y,k'),
\ \ k'>0.3\,{\rm GeV/c}
\eeq
again at $\bar{Y}=4,8,16$. One clearly observes that below a certain point all
the momentum dependence is reduced to the trivial factor $1/k^2$, implying that
the integral
in Eq. ( 4 ) is independent of the momentum. This point roughly coincides
with the saturation momentum $Q_s(y,\beta)$.
However one should have in mind that for a given nucleus the
value of the saturation momentum varies depending on the nuclear
transverse density at distance $\beta$ from the center of
the nucleus. For $A\geq 9$ and values of
$\beta$ inside the nucleus we find
$Q_s \sim 20\div 200 $ GeV/c at $\bar{Y}/2=4$  and
$Q_s\sim 1.5.\, 10^5\div 1.5.\, 10^6$ at $\bar{Y}/2=8$. Comparing with Fig. 4
we find that at $k<Q_s$ the integral factor in ( 4 ) is practically a constant.
For $k>Q_s$ it rapidly falls.  At large $k>>Q_s$ the inclusive cross-sections
at the center $y=Y/2$ are found to behave
like $1/k^p(y)$ with power $p(y)$ diminishing with energy. From our calculations
we approximately find that $p(y)=3.3,\ ,3.0$ and 2.7at $\bar{y}=2,\,4$ and 8
respectively. At infinite energies $p$ seems to tend to 2 in correspondence
to the fact that $Q_s\to\infty$

In Fig. 5   we illustrate the $A$-dependence showing ratios
\beq
R^{part}_{A}= \frac{9I_{A}(y,k)}{AI_{9}(y,k)}
\eeq
with inclusive
cross-sections scaled  by $A$, at $y=Y/2$ and $\bar{Y}=4,8$ and 16
(from top to bottom). One clearly sees that whereas at relatively
small momenta the inclusive cross-sections are proportional to $A$,
that is to the number of participants, at larger momenta they grow
with $A$ faster, however  noticeably slowlier than the number of
collisions, roughly as $A^{1.1}$. The interval of momenta for which
$I_A\propto A$ can also be related to the value of the  saturation
momentum $Q_s(y,\beta)$. Inspection of
 Fig. 5 shows
that the distributions are proportional to $A$ at values of $k$ smaller
or in the vicinity of the value of the saturation momentum. Since $Q_s(y,\beta)$
grows with energy, one may conjecture that at infinite energies all the spectrum
will be proportional to $A$.

Passing to the determination of multiplicities one has to observe
certain care because of the properties of the perturbative QCD
solution in the leading approximation embodied in Eqs. (4) and (7).
As follows from ( 4 ) the inclusive cross-section blows up at
$k^2\to 0$ independently of rapidity $y$. So the corresponding
total multiplicity diverges logarithmically. However, the physical
sense has only emission of jets with high enough transverse momenta.
Thus one has to cut the spectrum from below by some $k_{min}$ which
separates the spectrum of jets proper from soft gluons which are not
related to jets. Inevitably the multiplicity of thus defined jets
depends on the  chosen value of $k_{min}$. We have chosen
$k_{min}=2$ GeV/c. At all energies the multiplicities at the center
$\mu_A(y=Y/2)$ were found to be approximately proportional to $A$.
The ratios $\mu_A(y=Y/2)/A$ are presented in Fig. 6. One observes
that their values are quite high and grow very fast with energies.
This is not surprising, considering a very high value of the pomeron
intercept in the lowest order BFKL model. The bulk of the
multiplicity comes from jets with relatively high momenta. To
illustrate this point in Fig. 7 we show   the dependence of the
central multiplicity on $k_{min}$ for $A=9$ and at $\bar{Y}=8$ in
the interval $k_{min}=0.3\div 16$ GeV/c. One observes that it goes
down very slowly, indicating that it is the high momentum tail of
the distribution which matters.

Finally we pass to the cross-sections obtained with the KT formula (
6 ). In Fig. 8 we show the ratios of these cross-sections to the
ones defined by the AGK rules, Eq. ( 4 ), for $y=Y/2$ and
$\bar{Y}=4,8$ and 16. These ratios turn to unity at $k$ in the
vicinity and above $Q_s$, as discussed in the end of the preceding
section. Below $Q_s$ the ratios are approximately equal to $0.8\div
0.9$ with little dependence on $A$ and $Y$. Some dependence which is
left can be explained by the fact that for very peripheral parts of
the nucleus the scaling regime can only be reached at rapidities
well above the considered ones. Due to this very simple relation
between the two cross-sections, all conclusions about the $A$
-dependence drawn for the AGK cross-section ( 4 ) remain valid also
for the KT cross-section ( 6 ).

 \section{Conclusions}
We have calculated the inclusive cross-sections for gluon production
at mid-rapidity in nucleus-nucleus central collisions in the
perturbative QCD approach with a large number of colors. Realistic
nuclear densities were employed to account for the peripheral parts
of the nuclei, whose contribution rapidly grows with energy due to
smallness of unitarizing non-linear effects.
 The form of
the cross-sections is found to be determined by the value of the
saturation momentum $Q_s$, which depends on the rapidity and nuclear
density. At momenta much lower than $Q_s$ the spectrum is
proportional to $1/k^2$. Its $A$ dependence is close to linear. At
momenta much higher than $Q_s$ the spectrum is found to fall
approximately as $1/k^{2.7\div 3.3}$ with the $A$-dependence as
$\sim A^{1.1}$. The multiplicities at mid-rapidity are found to be
proportional to $A$ with a good precision. They grow with energy
very fast which is related  to a fast growth of the saturation
momentum.

We also compared two different forms for the inclusive
cross-section, which follow from the AGK rules or the dipole
picture. The difference between their predictions was found to be
absent for values of momenta larger than $Q_s$. At momenta smaller
than $Q_s$ the difference reduces to a universal constant factor:
the dipole cross-sections are just $0.8\div 0.9$ of the AGK cross-sections.
With the growth of $Y$ and/or $A$ this factor slowly grows towards unity,
so that it is not excluded that at infinite $A$ and $Y$ the two cross-sections
( 4 ) and ( 6 ) totally coincide at all values of momenta.
All our conclusions about the energy, momentum and $A$ dependence
are equally valid for both forms of the inclusive cross-sections.

As mentioned in the Introduction a few more phenomenological studies
of the gluon production in nucleus-nucleus collisions were recently
made in the classical approximation to the colour-glass condensate
model ~\cite{KNV} and in the saturation model of ~\cite{KLN}. In
both studies quantum evolution was neglected, so that scaling with
the saturation momentum $Q_s$ was postulated rather than derived.
The saturation momentum thus appeared as an external parameter,
whose $A$ and $Y$ dependence were chosen on general grounds and
whose values were fitted to the experimental data at RHIC. In both
models the multiplicities turned out to be proportional to the
number of participants (modulo logarithmic dependence on $A$
different in the two approaches). This agrees with our results.
However the form of the inclusive distributions in momenta found in
~\cite{KNV} is  different from ours. Its behaviour both at small $k$
($\sim 1/\sqrt{k^2+m^2}$ with $m=0.0358 Q_s$) and at large
 $k$ ($\sim 1/k^4 $)
disagrees with the form of the spectrum we have found. For realistic
nuclei the spectrum
was calculated in ~\cite{KNV} only up to 6$\div$7 GeV/c, so it is not
possible to see if any change in its $A$-behaviour will occur at
higher momenta.  But most of all, the value of the saturation
momentum and the speed of its growth with rapidity which we have
found from the QCD pomeron model with full quantum evolution are
much larger than the fitted values in both ~\cite{KNV} and
~\cite{KLN}. This is no wonder in view of a very high value of the
BFKL intercept in the leading approximation. From the
phenomenological point of view this is the main drawback of the BFKL
theory. To cure it one possibly has to include  higher orders
of the perturbation expansion and  the running coupling constant.
Although some work in this direction has been done for linear
evolution ~\cite{LFC},  no attempts to generalize this to non-linear
evolution in some rigour has been made yet.


\section{Acknowledgements}
The author is deeply indebted to Yu.Kovchegov for a constructive
discussion and valuable comments and to N.Armesto and B.Vlahovic for their
interest in this work. This work has been supported by
a NATO Grant PST.CLG.980287.

\newpage

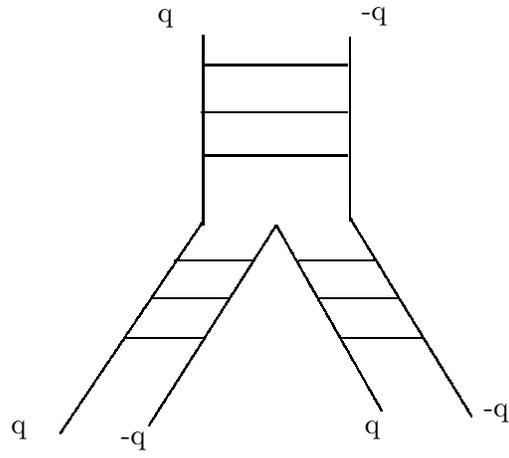
\begin{figure} [ht]
\unitlength 1mm 
\linethickness{0.4pt}
\ifx\plotpoint\undefined\newsavebox{\plotpoint}\fi 
\begin{picture}(100.25,103.75)(0,0)
\put(61.25,101.25){\line(0,-1){25.25}}
\put(80.75,101){\line(0,-1){24.5}}
\multiput(61.25,76.5)(-.033687943,-.050088652){564}{\line(0,-1){.050088652}}
\put(42.25,48.25){\line(0,1){0}}
\put(80.75,76.75){\line(1,0){.25}}
\multiput(81,76.75)(-.03125,.03125){8}{\line(0,1){.03125}}
\multiput(80.75,77)(.033673469,-.055102041){245}{\line(0,-1){.055102041}}
\multiput(89,63.5)(.03361345,-.05462185){238}{\line(0,-1){.05462185}}
\put(71,75.75){\line(0,1){.25}}
\multiput(71,76)(-.033730159,-.053075397){504}{\line(0,-1){.053075397}}
\multiput(71,76)(.03373494,-.059638554){415}{\line(0,-1){.059638554}}
\put(61.25,97.25){\line(1,0){19.25}}
\put(61,91){\line(1,0){19.5}}
\put(61.25,85.25){\line(1,0){19.25}}
\put(57.5,71.25){\line(1,0){10.25}}
\put(59.25,65){\line(0,1){0}}
\put(54.25,66.25){\line(1,0){10.5}}
\put(51,61){\line(1,0){10.25}}
\put(74,71.25){\line(1,0){10}}
\put(76.5,66.25){\line(1,0){10.5}}
\put(79.5,61){\line(1,0){11}}
\put(56.25,103.5){\makebox(0,0)[cc]{q}}
\put(36.75,48.75){\makebox(0,0)[cc]{q}}
\put(84,103.75){\makebox(0,0)[cc]{-q}}
\put(100.25,51){\makebox(0,0)[cc]{-q}}
\put(52,47.25){\makebox(0,0)[cc]{-q}}
\put(83.75,48.75){\makebox(0,0)[cc]{q}}
\end{picture}
\vspace*{-2 cm}

\caption{Interaction of three BFKL pomerons at the
spliiting vertex} \label{Fig1}
\end{figure}
\begin{figure}
\unitlength 1mm 
\linethickness{0.4pt}
\ifx\plotpoint\undefined\newsavebox{\plotpoint}\fi 
\begin{picture}(175.25,101.25)(0,0)
\multiput(22,97.25)(.033713693,-.04719917){482}{\line(0,-1){.04719917}}
\multiput(38,75)(.033723022,.039568345){556}{\line(0,1){.039568345}}
\put(37.5,75){\line(0,-1){21.25}}
\multiput(37.5,53.75)(-.033713693,-.034751037){482}{\line(0,-1){.034751037}}
\multiput(37.5,53.5)(.03372093,-.036046512){430}{\line(0,-1){.036046512}}
\multiput(38.5,97)(-.033636364,-.036363636){275}{\line(0,-1){.036363636}}
\multiput(43.25,47.5)(-.033737024,-.033737024){289}{\line(0,-1){.033737024}}
\multiput(81.75,99.5)(.033713693,-.04719917){482}{\line(0,-1){.04719917}}
\multiput(97.75,77.25)(.033723022,.039568345){556}{\line(0,1){.039568345}}
\put(97.25,77.25){\line(0,-1){21.25}}
\multiput(97.25,56)(-.033713693,-.034751037){482}{\line(0,-1){.034751037}}
\multiput(97.25,55.75)(.03372093,-.036046512){430}{\line(0,-1){.036046512}}
\multiput(98.25,99.25)(-.033636364,-.036363636){275}{\line(0,-1){.036363636}}
\multiput(103,49.75)(-.033737024,-.033737024){289}{\line(0,-1){.033737024}}
\multiput(140.5,101.25)(.033713693,-.04719917){482}{\line(0,-1){.04719917}}
\multiput(156.5,79)(.033723022,.039568345){556}{\line(0,1){.039568345}}
\put(156,79){\line(0,-1){21.25}}
\multiput(156,57.75)(-.033713693,-.034751037){482}{\line(0,-1){.034751037}}
\multiput(156,57.5)(.03372093,-.036046512){430}{\line(0,-1){.036046512}}
\multiput(157,101)(-.033636364,-.036363636){275}{\line(0,-1){.036363636}}
\multiput(161.75,51.5)(-.033737024,-.033737024){289}{\line(0,-1){.033737024}}
\put(28.75,67){\line(1,0){18}}
\multiput(85.5,77)(2.875,.03125){8}{\line(1,0){2.875}}
\put(108.5,77.25){\line(0,1){0}}
\put(148.25,57.5){\line(1,0){15.75}}
\put(36,30){\makebox(0,0)[cc]{a}}
\put(95.25,32.5){\makebox(0,0)[cc]{b}}
\put(157.25,33.75){\makebox(0,0)[cc]{c}}
\end{picture}
\caption{Typical diagrams for the inclusive cross-section in nucleus-nucleus
collisions.}
\label{Fig2}
\end{figure}
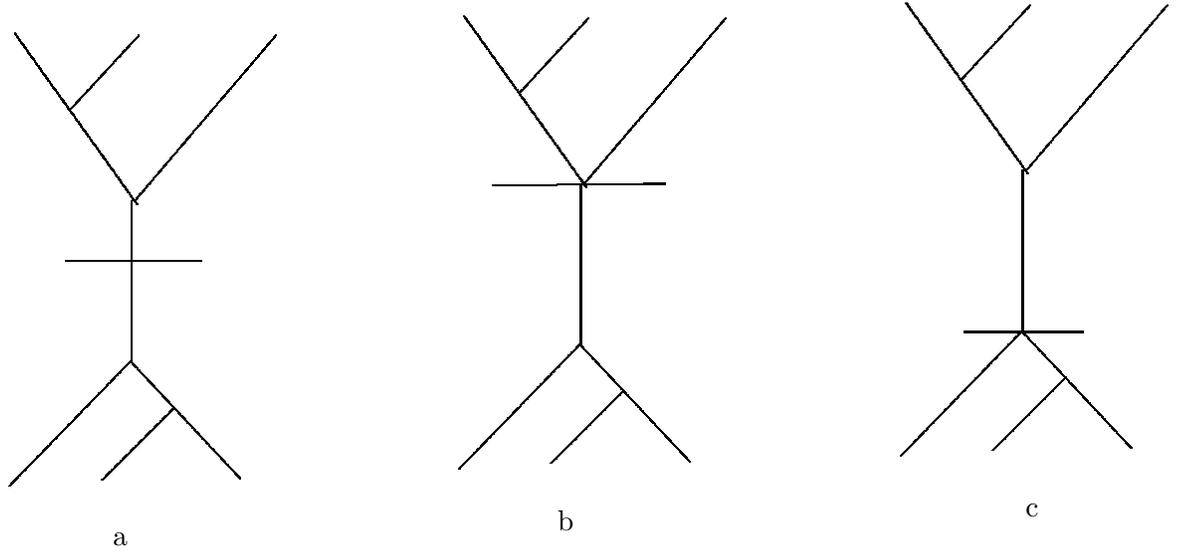
\begin{figure}[ht]
\epsfxsize 4in
\centerline{\epsfbox{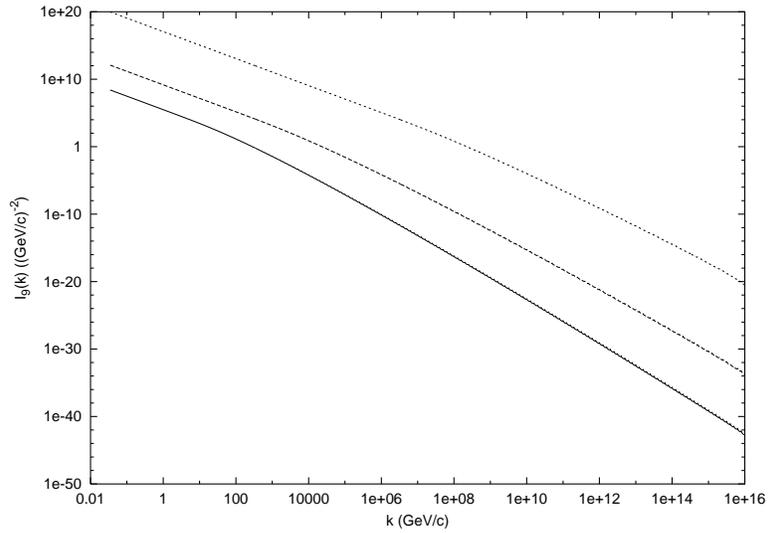}}
\caption{Inclusive cross-sections $I_9(y,k)$
at $y=\frac{Y}{2}$ .
Curves from top to bottom at small $k$ correspond to scaled overall
rapidities $\bar{Y}=4,8,16$.}
\label{Fig3}
\end{figure}
\begin{figure}[ht]
\epsfxsize 4in
\centerline{\epsfbox{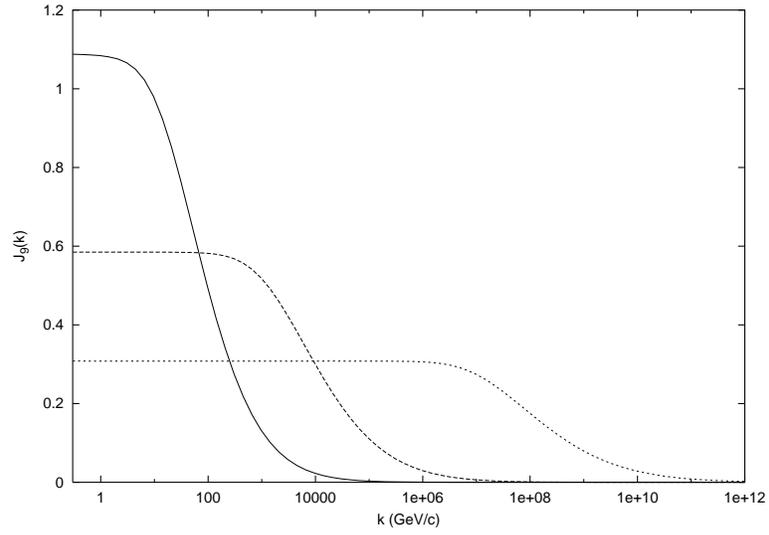}}
\caption{Normalized distributions $J_9(y,k)$ (Eq. (25))
at $y=\frac{Y}{2}$ .
Curves from top to bottom at small $k$ correspond to scaled overall
rapidities $\bar{Y}=4,8,16$.}
\label{Fig4}
\end{figure}
%
%
%
%
\begin{figure}[ht]
\epsfxsize 4in
\centerline{\epsfbox{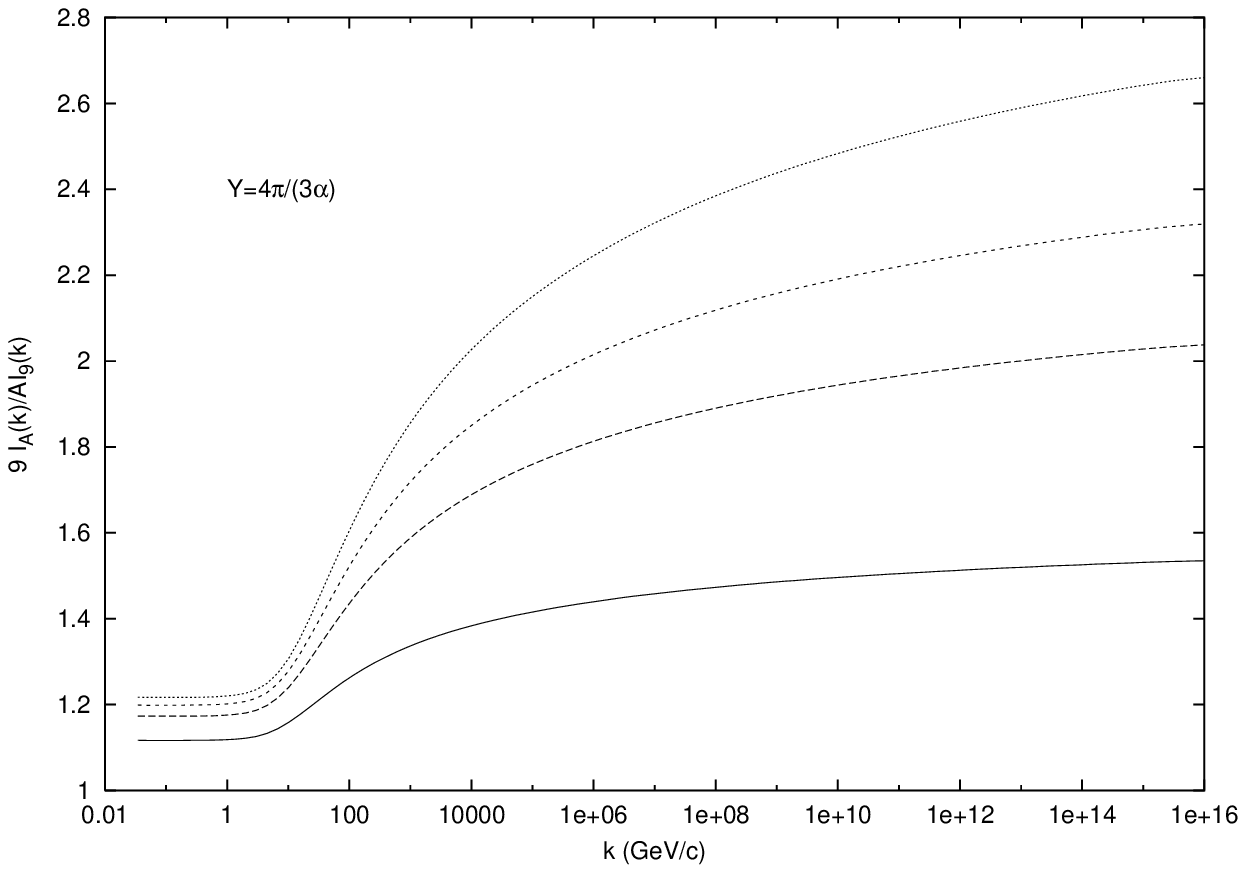}}
%
\epsfxsize 4in
\centerline{\epsfbox{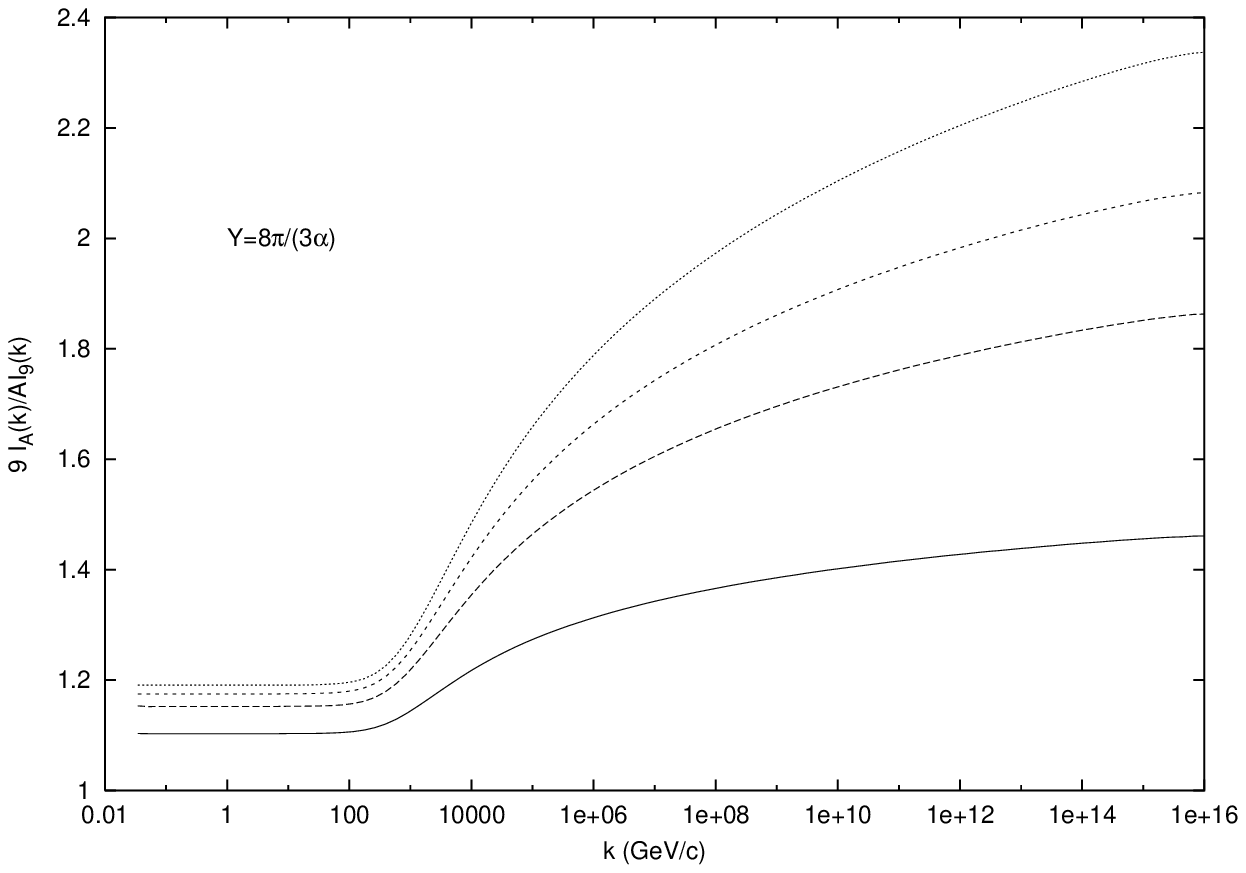}}
%
\epsfxsize 4in
\centerline{\epsfbox{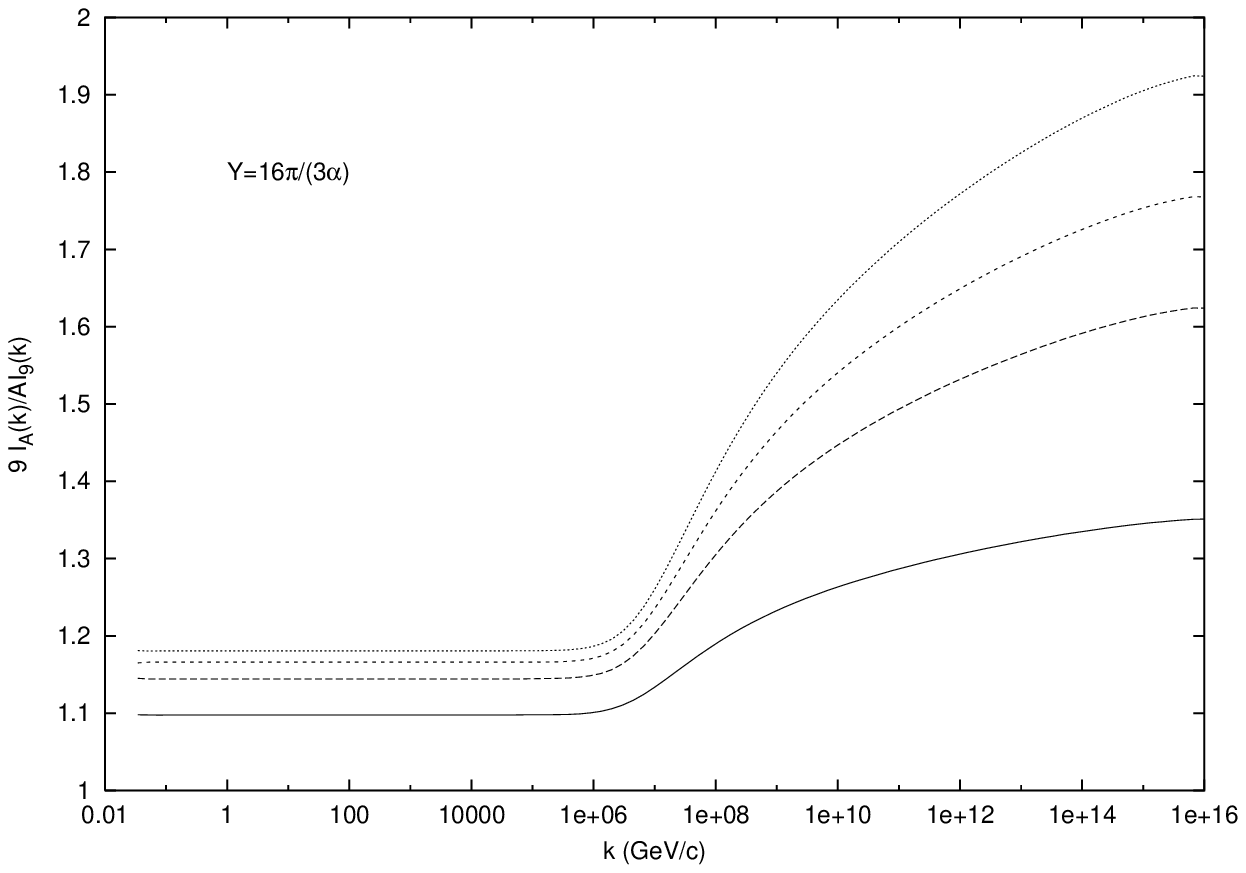}}
\caption{From top to bottom: A-dependence of momentum distributions,
scaled with the number of participants, at $\bar{Y}=4, 8,16$.
Curves from bottom to top show ratios $9I_A(y,k)/AI_9(y,k)$
at center rapidity ($y=\frac{Y}{2}$) for $A=9,27,64,108$ and 180.}
\label{Fig5}
\end{figure}
\begin{figure}[ht]
\epsfxsize 4in
\centerline{\epsfbox{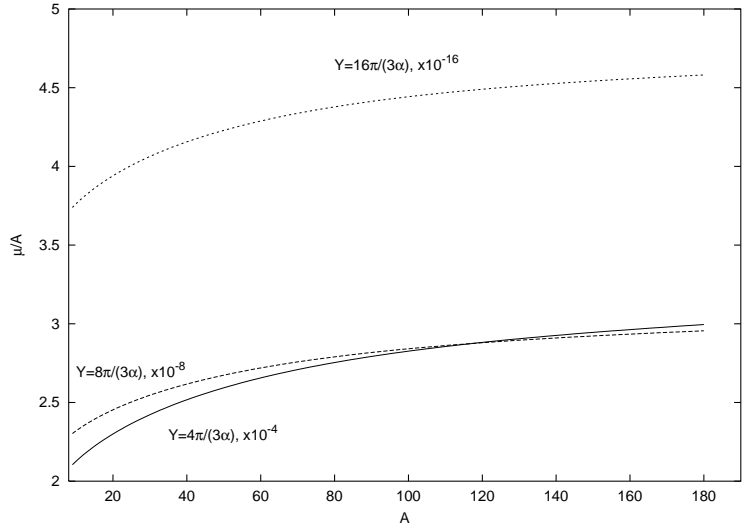}}
\caption{From top to bottom: A-dependence of multiplicities,
scaled with the number of participants, at $\bar{Y}=4, 8,16$.
Curves from bottom to top show  $\mu_A(y)/A$
at center rapidity ($y=\frac{Y}{2}$) for $A=9,27,64,108$ and 180.}
\label{Fig6}
\end{figure}
\begin{figure}[ht]
\epsfxsize 4in
\centerline{\epsfbox{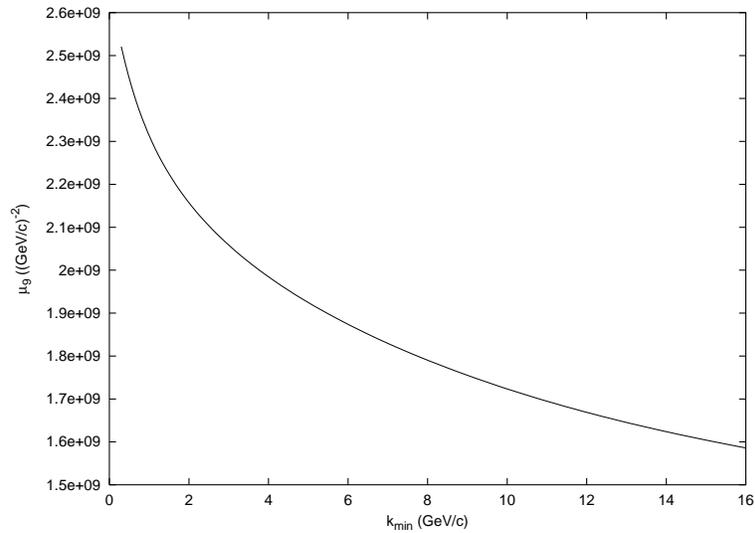}}
\caption{The dependence of multiplicity for $A=9$
at center rapidity and $\bar{Y}=8$ on the infrared cut
$k_{min}$}.
\label{Fig7}
\end{figure}
\begin{figure}[ht]
\epsfxsize 4in
\centerline{\epsfbox{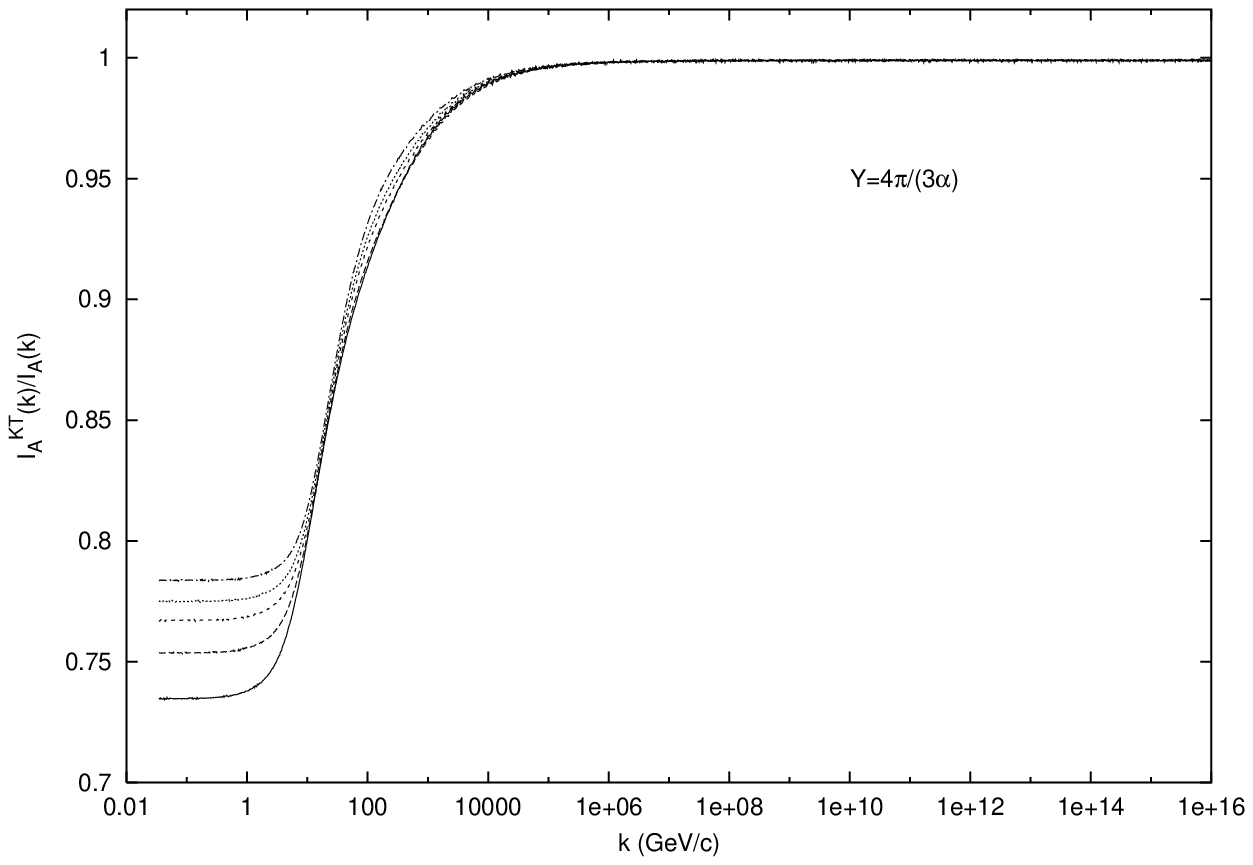}}
\epsfxsize 4in
\centerline{\epsfbox{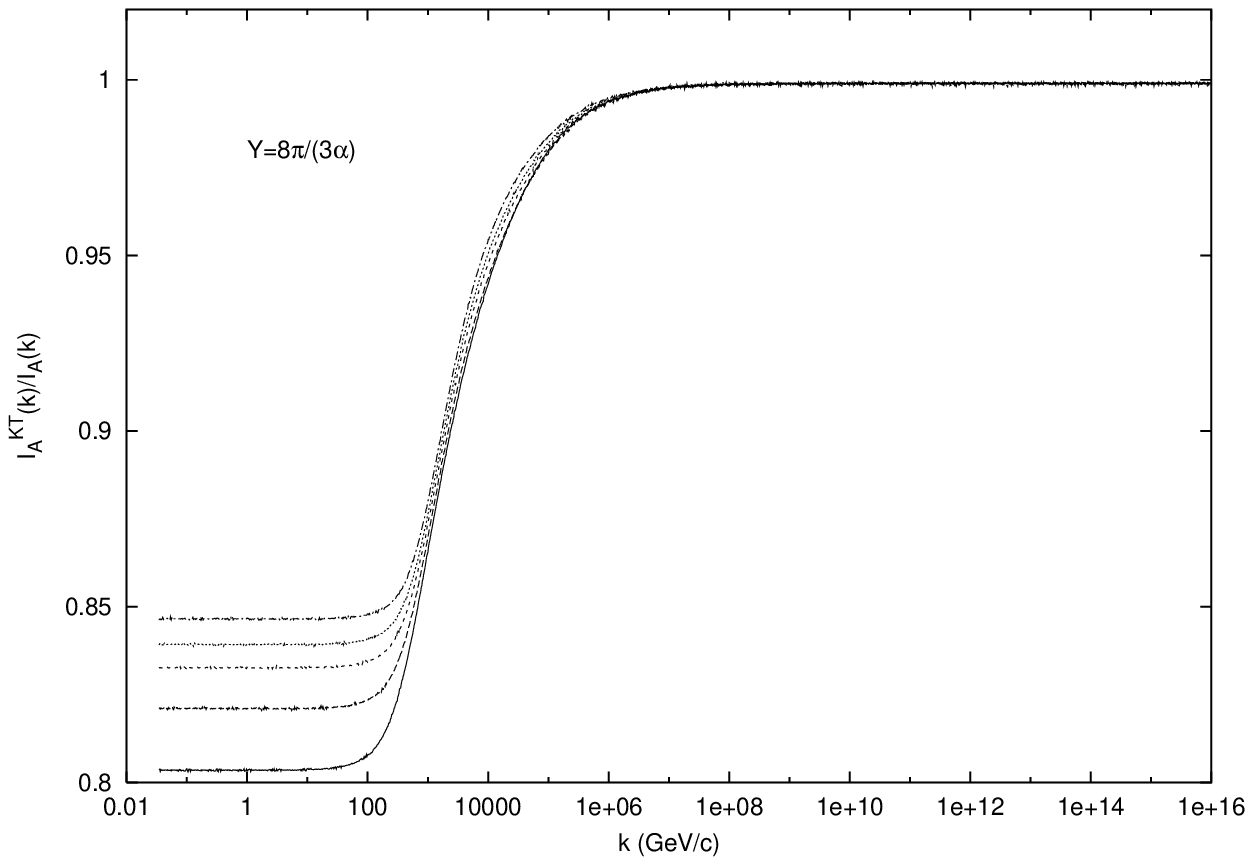}}
\epsfxsize 4in
\centerline{\epsfbox{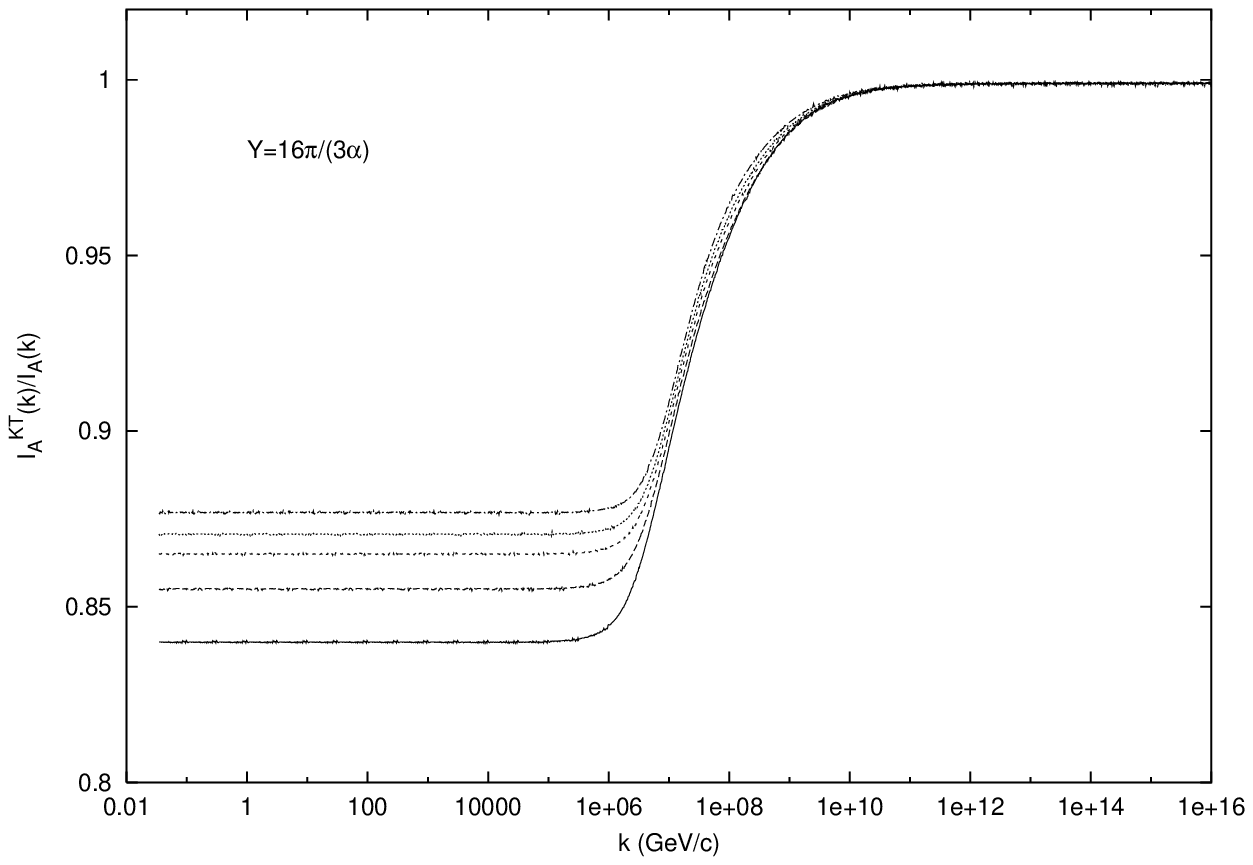}}
\caption{Ratios of the KT inclusive cross-sections, Eq. ( 6 ), to the
ones found from the AGK rules, Eq. (4) at
center rapidity ($y=Y/2$) and $\bar{Y}=4, 8,16$.
Curves from bottom to top refer to $A=9, 27, 64, 108$ and 180 }
\label{Fig8}
\end{figure}
\end{document}